\documentclass[prl,preprintnumbers,twocolumn,floatfix,nofootinbib,a4paper,superscriptaddress]{revtex4-2}
\usepackage{color}
\usepackage{calc}
\usepackage{graphicx}
\usepackage{amssymb,amsmath}
\usepackage{tensor}
\usepackage{bm}
\usepackage{float}
\usepackage{microtype}
\usepackage{booktabs}
\usepackage{times}
\usepackage[varg]{txfonts}
\usepackage[colorlinks, pdfborder={0 0 0}]{hyperref}
\usepackage{subfigure}

\hypersetup{citecolor=blue}

\allowdisplaybreaks
\usepackage[utf8]{inputenc}
\usepackage{ulem}
\usepackage{multirow}
\usepackage{array}    
\usepackage{ulem,cancel}
\usepackage{upgreek} 

\usepackage{comment}
\usepackage[nolist]{acronym}
\usepackage{tikz}
\usepackage{textcomp}
\usepackage[export]{adjustbox}
\usepackage{url}

\usepackage{dcolumn}
\usepackage{bm}
\usepackage[]{xcolor}

\newcommand{\wt}{\tilde}
\newcommand{\p}{\partial}
\newcommand{\ti}{\textit}

\newcommand{\const}{\text{const}}  
\newcommand{\etad}{\eta_{\rm d}} 
\newcommand\vect[1]{\mathbf #1}
\newcommand\mean[1]{\langle #1\rangle}

\newcommand{\dd}{\mathrm{d}}        
\newcommand\e{\mathrm{e}} 
\newcommand{\ii}{\mathrm{i}} 
\newcommand\oderiv[2]{\frac{\dd#1}{\dd#2}}
\newcommand{\Rey}{\text{Re}}  
\newcommand{\Rm}{R_\text{m}}  
\newcommand{\Pm}{\text{Pm}}  

\newcommand{\fr}{_*}  
\newcommand{\nl}{_\text{nl}}  
\newcommand{\tff}{t_{\rm ff}}

\newcommand{\cm}{\,{\rm cm}}     
\newcommand{\km}{\,\text{km}}    
\newcommand{\psc}{\,{\rm pc}}     

\newcommand{\g}{\,{\rm g}}      

\newcommand{\s}{\,{\rm s}}      
\newcommand{\yr}{\,{\rm yr}}    
\newcommand{\Myr}{\,{\rm Myr}} 
\newcommand{\Gyr}{\,{\rm Gyr}}  
\newcommand{\kms}{\km\s^{-1}}    

\newcommand{\G}{\,{\rm G}}      

\usetikzlibrary{arrows,shapes,trees,decorations.pathreplacing}
\usepackage{aas_macros}
\usepackage{diagbox}

\allowdisplaybreaks

\begin{document}


\title{Turbulent dynamos in a collapsing cloud}

\author{Muhammed Irshad P}
\email{muhammed.irshad@icts.res.in}
\author{Pallavi Bhat}
\affiliation{International Centre for Theoretical Sciences, Tata Institute of Fundamental Research, Bangalore 560089, India}

\author{Kandaswamy Subramanian}
\affiliation{Inter-University Centre for Astronomy and Astrophysics, Post Bag 4, Ganeshkhind, Pune 411007, India}
\affiliation{Department of Physics, Ashoka University, Rajiv Gandhi Education City, Rai, Sonipat 131029, India}

\author{Anvar Shukurov}
\affiliation{School of Mathematics, Statistics and Physics, Newcastle University, Newcastle upon Tyne, NE1 7RU, U.K.}



\begin{abstract}
	The amplification of magnetic fields is crucial for understanding the observed magnetization of stars and galaxies. Turbulent dynamo 
	is the primary mechanism responsible for 
	that but the understanding of its action in a collapsing environment is still rudimentary and relies on limited numerical experiments.
	We develop an analytical framework and perform numerical simulations to investigate the behavior of small-scale and large-scale dynamos in a collapsing turbulent cloud. 
	This approach is also applicable to expanding environments and facilitates the application of standard dynamo theory to evolving systems. Using a supercomoving formulation of the magnetohydrodynamic (MHD) equations, we demonstrate that dynamo action in a collapsing background leads to a \textit{super-exponential} growth of magnetic fields in time, significantly faster than the 
	exponential growth seen in stationary turbulence. The enhancement is mainly due to the increasing eddy turnover rate during the collapse, which boosts the instantaneous growth rate of the dynamo. We also show that the {scaling of} final saturated magnetic field strength {with density robustly} exceeds the expectation from considerations of pure flux-freezing{.}
	Apart from establishing a formal framework for 
	studying 
	magnetic field evolution in collapsing (or expanding) turbulent plasmas, these findings {suggest that during 
		star and galaxy formation} 
	magnetic fields can be{come} 
	dynamically relevant 
	much earlier than previously thought.
\end{abstract}

\maketitle



\textit{Introduction.---}
Coherent and random magnetic fields are ubiquitous in stars and galaxies \citep{reiners_observations_2012, hull_interferometric_2019,beck2015,Borlaff+2023}. The leading paradigm that has the potential to explain the origin of these magnetic fields is the turbulent dynamo theory \citep{BRANDENBURG20051, shukurovkandubook}. 
In a highly conducting turbulent 
plasma, the {\it small-scale dynamo} (SSD) amplifies  magnetic fluctuations on a typical time scale of order the turnover time of a turbulent eddy, whereas the {\it large-scale dynamo} (LSD) generates fields coherent on scales larger than the turbulent scale
on a longer time scale. In galaxies, the two time scales are of order $10\Myr$ and 5--$10\Gyr$, respectively. 

{M}agnetic fields coherent on scales of a few kiloparsecs appear to be present in galactic and protogalactic environments even at high redshifts \citep{bernet_strong_2008,farnes_faraday_2014, farnes_observed_2017, mao_detection_2017, chen_kiloparsec-scale_2024}. Their amplification on the corresponding time scale of a few Gyr requires a special explanation as suggested here.

{
	The early generation of magnetic fields can influence the stellar mass spectrum by suppressing fragmentation and enabling jets and outflows from accretion disks \cite{Robi2006, JoosAA2013, krumholz_role_2019, ray_jets_2021, hennebelle_influence_2022, refId0, Sadanari2024}. They also shape galaxy evolution and drive feedback through winds and fountains \cite{pudritz_magnetic_2012, ntormousi_magnetic_2018, sanati_dwarf_2024}.
}

The formation of stars and galaxies starts with the collapse of a gas cloud driven by gravitational, thermal and other instabilities leading to turbulence \citep{O’Shea_2007, Greif_Firstgalaxy2008, Yoshida_2008, Kl_Hen2010Accretion, sur_generation_2010, Lee_2015, Hu_2021, archana_IND2024}{, which}
can be amplified  during the gravitational contraction \citep{Higashi_2021, Higashi_2022, Vázquez-Semadeni_1998, birnboim2017, Robertson_2012, Ankush_Fed20, Hennebelle2021, Xu_2020}.
This presents an intriguing possibility of the dynamo process receiving a boost from the collapse leading to an accelerated growth of magnetic fields. Previous analytical studies \citep{Mckee1, Xu_2020NLcol, Axel22,  higashi_amplification_2024}  
{do not explore this possibility.}
Also, accelerated growth in numerical simulations can be difficult to detect without the right parameter regime 
and 
systematic comparisons across parameter variations 
\citep{Federrath_2011, sur2012, McKee2}.

In this letter, we develop an analytical framework for dynamo action in a collapsing cloud, {demonstrating} super-exponential growth of the magnetic field strength. Nonlinear regimes of magnetic field evolution in such environments are explored using numerical simulations. 


\textit{Collapse of a homogeneous cloud.---}
Consider a turbulent cloud undergoing homologous collapse. The position of the fluid element and its density are given by, 
\begin{equation}\label{eq:smhd2}
\vect{r} = a(t) \vect{r}_\text{c},\quad \rho(t) = \rho_\text{c}/a^3(t), \quad 0 < a(t) \le 1,
\end{equation}
where $a(t)$ is the scale factor determining the size, $\vect{r}_\text{c}$ and $\rho_\text{c}$ are the comoving position and density, respectively. 

Even in the comoving coordinates, the magnetohydrodynamics (MHD) equations are still intricate because of the contribution from the background collapse to the total velocity of fluid element given by {
	$\vect{u} = \vect{v} + H\vect{r}$.} Here $\vect{v}$ is the peculiar velocity and $H\vect{r}$ is the velocity of the collapsing background {with the Hubble parameter
	$H = \dd\ln a/\dd t$.} 
Following \citep{shapiro97, Shandarin1980}, we introduce the so-called 
supercomoving variables denoted here with tilde,
\begin{equation}\label{eq1:mhd_sum}
\begin{split}
\tilde{\vect{r}} &= \vect{r}/a,\quad \dd \tilde{t} = \dd t/{a^{2}},\quad\wt{\rho}(\wt t) =a^{3} \rho(t),\\
\wt{\vect{v}}(\wt t) &= a\vect{v}(t), \quad \wt{p}(\wt t) = a^{5}p(t),\quad \wt{\vect{B}}(\wt t) = a^{2}\vect{B}(t),
\end{split}
\end{equation}
where $p$ is the pressure and $\vect{B}$ is the magnetic field. This results in a simpler structure of the governing equations. 
Note that we use the supercomoving time $\wt{t}$ instead of $t$ in the analysis and $\wt{\vect{v}}$ accounts for the evolution of peculiar velocity 
as $1/a$ in a homologous collapse \citep{Peebles}. Similarly, $\wt{\vect{B}}$ is defined so that the magnetic field amplification via flux freezing, in proportion to $1/a^2$, due to the overall collapse is factored out. {In physical variables, this represents an adiabatically collapsing sphere, evident from pressure and density scaling.}

For an incompressible comoving flow, the MHD equations in supercomoving coordinates 
follow as  
\begin{align}
\wt{\nabla}\cdot\wt{\vect{v}} &= 0, \label{eq2a:mhd_sum} 
\\
\frac{\p \wt{\vect{v}}}{\p \wt{t}} + (\wt{\vect{v}}\cdot\wt{\nabla})\wt{\vect{v}}
&= - \frac{\wt{\nabla} \wt{p}}{\wt{\rho}} +\wt{a}\frac{(\wt{\nabla}\times\wt{\vect{B}})\times\wt{\vect{B}}}{4\pi\wt{\rho}} + \nu \wt{\nabla}^2 \wt{\vect{v}},   \label{eq2b:mhd_sum}
\\
\frac{\p \wt{\vect{B}}}{\p \wt{t}} &= \wt{\nabla}\times(\wt{\vect{v}}\times\wt{\vect{B}})+\eta\wt{\nabla}^{2}\wt{\vect{B}},\quad \wt{\nabla}\cdot\wt{\vect{B}} = 0.\label{eq2c:mhd_sum}
\end{align}
The gravitational term 
is balanced by the acceleration of the collapse and, thus, does not appear in the momentum equation.
{This choice of incompressible supercomoving flow is consistent with simulations showing that solenoidal turbulence dominates in collapsing primordial cores, being enhanced by the spin-up of turbulent vortices \citep{Higashi_2021}}.
These equations have the same form as the MHD equations in a stationary background except for the factor {$\tilde{a}(\tilde t) \equiv a(t)$} in the Lorentz force.
The major advantage of these variables is that the induction equation retains its form. This helps to extend the
standard kinematic dynamo theory to a homologous collapsing background.

The form of $a(t)$ is governed by the Friedmann equation 
{
	${\Ddot{a}}/{a} = - 4\pi G \rho(t) / 3,$ 
}
where the dot denotes the time derivative, which
can be integrated with the free-fall initial conditions $a(0) = 1$, $\dot{a}(0) = 0$: 
\begin{equation}\label{eq:fried2}
\arctan\left(\sqrt{(1-a)/a}\right) 
+ \sqrt{a(1-a)}= {s}t,
\end{equation}
where ${s}=\sqrt{8\pi G \wt{\rho}/3}$. In the limit $a\to 0$, $t$ tends to the free-fall time, $t_{\rm ff} = {\pi}/{(2{s})} = \left[{3\pi}/{(32 G \wt\rho)}\right]^{1/2},$ which is inversely proportional to the initial density of the plasma.
Taking the differential of Eq.~\eqref{eq:fried2}, we have $\dd\wt{t} = \dd t/a^2=-{s}^{-1} a^{-3/2}(1-a)^{-1/2}\,\dd a$, which {gives the relation,}
\begin{equation}\label{eq:fried5}
\wt{t} = \frac{2}{{s}}\sqrt{\frac{1-a}{a}} \equiv f(t),
\end{equation}
where the constant of integration follows from the requirement that $\tilde t=0$ at $a=1$ and $t=0$.
Inverting Eq.~\eqref{eq:fried5} gives the scale factor in terms of $\tilde t$, 
\begin{equation}\label{eq:sim6}
\wt{a}(\wt{t}) = \frac{1}{1 + {{s}^{2}\wt{t}^{2}}/{4}}.
\end{equation}
Note that $a\to 0,\: t\to t_{\rm ff}$ in the limit $\tilde t \to \infty$. Thus, we cannot track the collapse up to the singularity in terms of the supercomoving time. A fit 
for the scale factor (accurate within 2\%) given by $a(t) = \left(1 - {t^{2}}/{t_{\rm ff}^{2}}\right)^{2/3}$, shows that initially the collapse is slow but it accelerates to reach a singularity at $t=t_{\rm ff}$ \cite{Girichidis}. 
The gas pressure stops the collapse before a singular state is reached as the cloud becomes sufficiently inhomogeneous and spherical symmetry breaks \citep{V-sPB-PGZ19}.
Therefore, in the simulations discussed below, the collapse is only extended to a certain earlier time $t\fr$, after which magnetic field continues evolving against a stationary background.


\textit{Turbulent dynamos in a collapsing background.---}In a stationary environment, the coefficients of the induction equation are independent of time, so dynamo action results in an exponentially fast amplification of an initial (weak) magnetic field at the expense of the kinetic energy of the plasma \citep{Kazantsev, Kandu1997GR}. The exponential growth continues until the Lorentz force becomes comparable to the forces driving the plasma flow; after that, the exponential growth saturates. Since the induction equation, Eq.~\eqref{eq2c:mhd_sum} in the supercomoving variables has the same form as in a stationary background, the dynamo action causes an exponential amplification  of $\wt B$ in terms of the time variable $\wt t$,
{$
	\wt{B} \propto \exp{\left(\wt\gamma \wt t\right)}.
	$}
In a dynamo-passive system {(when the flow does not support dynamo action)}, $\wt\gamma<0$ because of the magnetic diffusion, but the dynamo action results in $\wt\gamma>0$. Suitable expressions for the growth rate $\wt\gamma$ of the turbulent dynamos (SSD and LSD) can be found in {\citep{SM, Bhat2014, Gopalakrishnan2024}}.
Using Eqs~\eqref{eq1:mhd_sum} and \eqref{eq:fried5} 
we obtain
the corresponding magnetic field strength in the rest frame,
\begin{equation}\label{eq:kcoldynamo3}
B \propto \frac{1}{a^2(t)}\exp\left(\int_0^t \frac{\wt{\gamma}\,\dd t'}{a^2(t')} \right)= \frac{1}{a^{2}(t)}\,\e^{\tilde{\gamma} f(t)},
\end{equation}
where the factor $1/a^2$ is due to the 
overall collapse and 
$\wt\gamma=\text{const}$ for both SSD and LSD. 
Instead of a constant growth rate in a stationary environment, the dynamo action in a  collapsing background produces a growth rate increasing with time because of the factor $f(t)$ in the exponent.
We refer to this regime as a
\ti{super-exponential} growth of the magnetic field. 
{This is a robust regime 
	{which persists} even 
	{if the} supercomoving turbulence in a collapsing cloud {is decaying} \cite{SM}.}

By the end of the kinematic regime,  the Lorentz force affects the plasma flow and the magnetic field strength reaches a stationary saturation level. The magnetic energy density becomes comparable to the turbulent energy density,
\begin{align}\label{eq:kcoldynamo4}
\wt{a}  {\wt{B}^2}/{8\pi} \simeq \tfrac{1}{2}\wt\rho \wt{v}^2
\end{align}
in terms of the rms values of $\wt{B}$ and $\wt{\rho}\wt{v}^2$, 
where the factor $\wt{a}$ is due to the coefficient of the Lorentz force in Eq.~\eqref{eq2b:mhd_sum}.
In terms of the physical saturated magnetic field strength, assuming $\wt\rho \wt{v}^2=\text{const}$, this implies {(see \citep{SM} for the decaying case)}
\begin{align}\label{eq:kcoldynamo5}
B \propto a^{-5/2} \propto \rho^{5/6}.
\end{align}
{This magnetic field strength significantly exceeds that due to pure flux freezing alone which leads to
	$B\propto \rho^{2/3}$, and/or that from considering dynamo in a stationary background.}


\textit{Kinematic dynamos.---}To verify and refine the analytical arguments, we solve
Eqs~\eqref{eq2a:mhd_sum}--\eqref{eq2c:mhd_sum} {and \eqref{eq:sim6}} numerically, with the induction equation written in terms of the vector potential.

A random flow 
at an energy-range scale $\wt{l}_0$ 
is driven by a random force added on the right-hand side of the momentum equation, Eq.~\eqref{eq2b:mhd_sum}.
We use the forcing function \cite{Candelaresi_2013} from the \textsc{Pencil Code} \cite{PCC21},
rewritten for the 
supercomoving coordinates
(see \citep{SM}). 
The magnitude of the force driving the random flow is independent of the scale while the forcing wavenumbers $\wt{\vect{k}}$ are random in direction and have 
magnitudes from the uniform distribution in the ranges $1\leq \wt{k}\leq 3$ (the average $\wt{k}_0=2\pi/\wt{l}_0=2$) 
for the SSD simulations and $3\leq \wt{k}\leq 5$ ($\wt{k}_0=4$) for the LSD.
{We note} that the {amplitude of the forcing increases and}
the forcing scale 
decreases with time in the {physical variables, which also allows for the additional driving by the gravitational contraction}
\cite{sur_generation_2010}. 

The driving force is mirror-symmetric in the SSD simulations and helical when the LSD is considered. The resulting LSD is of the $\alpha^2$ type because of the lack of differential rotation \cite{Sokoloff}. In differentially rotating clouds, the LSD can be significantly more efficient \cite{shukurovkandubook}. A random flow can be helical, and thus drive an LSD, in a rotating system. We neglect any deviation from spherical symmetry which may arise from the rotation.

We non-dimensionalize the equations with the forcing scale $\wt{l}_0$, the rms speed $\wt{v}_0$ (even in simulations without {forcing}), and {density} $\rho_0 = \wt{\rho}_0$, at $t=\wt{t}=0$. The unit time is $\wt{t}_0=\wt{l}_0/\wt{v}_0$ and the unit magnetic field is $\wt{B}_0=(4\pi\wt{\rho}_0 \wt{v}_0^2)^{1/2}$. The kinetic and magnetic Reynolds numbers, $\Rey=\wt{l}_0\wt{v}_0/\nu$
and $\Rm=\wt{l}_0\wt{v}_0/\eta$ are defined in the supercomoving variables and {remain constant}. 
For illustration, $\wt{t}_0\simeq10^8\yr$ and $\wt{B}_0\simeq 4\times10^{-7}\G$ for $\wt{l}_0=100\psc$, $\wt{v}_0=1\kms$, and $\wt{\rho}_0=10^{-24}\g\cm^{-3},$ parameters often used in the simulations of primordial star formation \cite{Mckee1}.

We used the publicly available code \textsc{Dedalus} \cite{Dedalus}, a
pseudospectral 
solver for partial differential equations. 
The simulations were
carried out in a periodic cubic box 
$(2\pi)^3$ in size,
with a resolution of $128^3$.
The initial vector potential is a random Gaussian noise with \text{a} strength $10^{-5}$. 

The kinetic
and magnetic Reynolds numbers 
used are 
$\Rey = \Rm = 415$ for the SSD simulations, 
while $\Rey = 180$ and $\Rm = 18$ for the LSD. The smaller $\Rm$ 
in the latter case 
support the LSD but not the SSD (which requires $\Rm\gtrsim100$ \cite{haugen}), 
so that we can study the two dynamo mechanisms separately. 

To reveal the effect of a collapsing background on dynamo action, we use three simulation runs for each of the SSD and LSD: (i) 
dynamo action 
in a stationary environment (referred to as the `standard' dynamo)
leading to an exponential magnetic field amplification, (ii) background collapse with $t_{\rm ff}=50$ without any random flow, thus no dynamo, and the magnetic field is only amplified due to magnetic flux freezing weakened by the magnetic diffusion, (iii) collapse with $t_{\rm ff}=50$ and driven random flow, so the field grows due to both dynamo action and the overall collapse of the cloud. The free-fall time chosen is shorter than the time at which the kinematic regime ends for the standard dynamo, $t\nl$  (which depends on the seed magnetic field strength and the dynamo amplification time scale).

\begin{figure}
	\centering    
	\includegraphics[width=3.4in, height=2.2in, keepaspectratio]{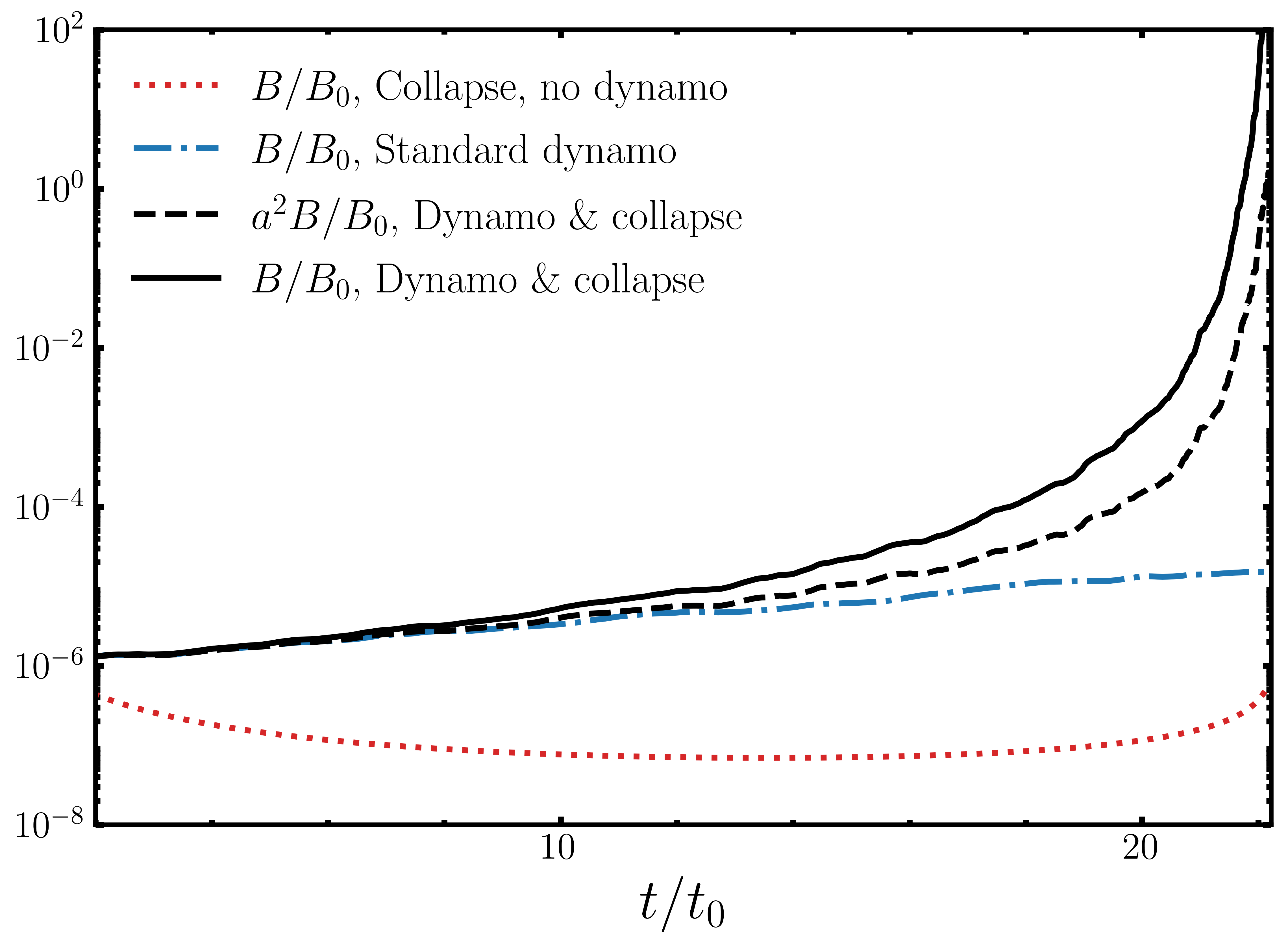}
	\caption{Comparison of magnetic field evolution in SSD {with $\Rey = \Rm = 415$} for different cases. Without collapse, there is exponential growth due to standard dynamo (dash-dotted).
		With collapse ($t_{\rm ff} = 50$) and without forcing (dotted), there is no dynamo; flux freezing competes with resistive diffusion. With both collapse and forcing the dynamo grows the magnetic field super-exponentially (dashed), which is further enhanced by the collapse (solid). The dashed curve {{isolates} the super-exponential growth by removing} the {factor} $1/a^2$ {arising from the} overall collapse{.} 
	} 
	\label{fig:compare_ssd}
\end{figure}

Figure~\ref{fig:compare_ssd} and {Fig.}~\ref{fig:compare_lsd} {in the end matter}  show the magnetic field evolution in the SSD and LSD, respectively, in the three scenarios. 
The collapse-induced compression of the magnetic field slows its resistive decay  
(dotted curves). When the magnetic diffusivity is relatively small ($\Rm = 415$, Fig.~\ref{fig:compare_ssd}), the compression becomes stronger than the decay at the final, accelerated stages of the collapse and the magnetic field strength slightly increases. However, the decay (which is also super-exponential) dominates at all times when the diffusivity is larger, $\Rm = 18$ (Fig.~\ref{fig:compare_lsd}). 
But, the magnetic field in a collapsing turbulent flow is both compressed by the overall collapse (approximately as $\propto a^{-2}$) and amplified by dynamo action.
The contribution of the dynamo is isolated in 
$a^2 B$ (dashed), and  $\log(a^2B)$ increases faster than {$\log(B)$ of the standard dynamo (dash-dotted)} showing its super-exponential growth. This confirms that the instantaneous growth rates of both SSD and LSD increase during the collapse. 
The solid lines in Figs~\ref{fig:compare_ssd} and \ref{fig:compare_lsd} show the evolution of the physical magnetic field strength confirming that is indeed strongly super-exponential.

\textit{Nonlinear dynamos.---}  
To investigate the nonlinear dynamos numerically, we stop the collapse at a certain time $t\fr$ well before the singular state is reached. After that, the scale factor remains constant at $a\fr$. If $t\fr$ occurs in the kinematic dynamo stage, the magnetic field continues growing exponentially (without acceleration) until the saturation. We also considered cases where the dynamo is already in the nonlinear stage at $t\fr$.
Thus, 
Eq.~\eqref{eq:sim6} 
is modified as
\begin{align}\label{eq:ssdafr1}
\wt{a}(\wt{t}) = 
\begin{cases}
(1 + {{s}^{2}\wt{t}^{2}}/{4})^{-1} & \text{if}\quad \wt t \le \wt{t}\fr, \\
(1 + {{s}^{2}\wt{t}\fr^{2}}/{4})^{-1} & \text{if}\quad \wt t > \wt{t}\fr.
\end{cases}
\end{align}
The resulting discontinuity in the time derivative of $\wt{a}$ leads to a discontinuity in the time variation of the Lorentz force. Therefore, the transition to a constant $\wt a$ was smoothed about $\wt{t}=\wt{t}\fr$ using a moving average.
To convert the supercomoving time $\wt{t}$ to the real time $t$, we numerically integrate 
$d\wt{t} =dt/a^2$ given in
Eq.~\eqref{eq1:mhd_sum}
using Eq.~\eqref{eq:ssdafr1}. We use $a\fr$ in the range 0.1--1 because smaller values require very long simulation runs.

The rate at which magnetic field is amplified in the kinematic dynamo stage depends on both the free-fall time $\tff$ (thus on the initial cloud density) and the time when the collapse stops, i.e., on $a\fr$ (if the dynamo is still kinematic at $t\fr$), whereas the steady-state magnetic field strength depends only on $a\fr$. We consider three cases {{where} $\tff$ {differently compares with} $t\nl$, the time at which standard dynamo becomes nonlinear:
} 
(i)~$t\nl > t_{\rm ff} = 50$, (ii)~$t\nl \sim t_{\rm ff} = 250~(\text{SSD}),\ 200~(\text{LSD})$, (iii)~$t\nl < t_{\rm ff} = 400${.} 
In all cases, we stop the collapse at two different 
times, $t\fr/t_\text{ff}=0.8$ and $0.95$ which correspond to different simulations with $a\fr\simeq 0.5$ and $0.2$, respectively.

\begin{figure}
	\centering    
	\includegraphics[width=3.4in, height=2.2in, keepaspectratio]{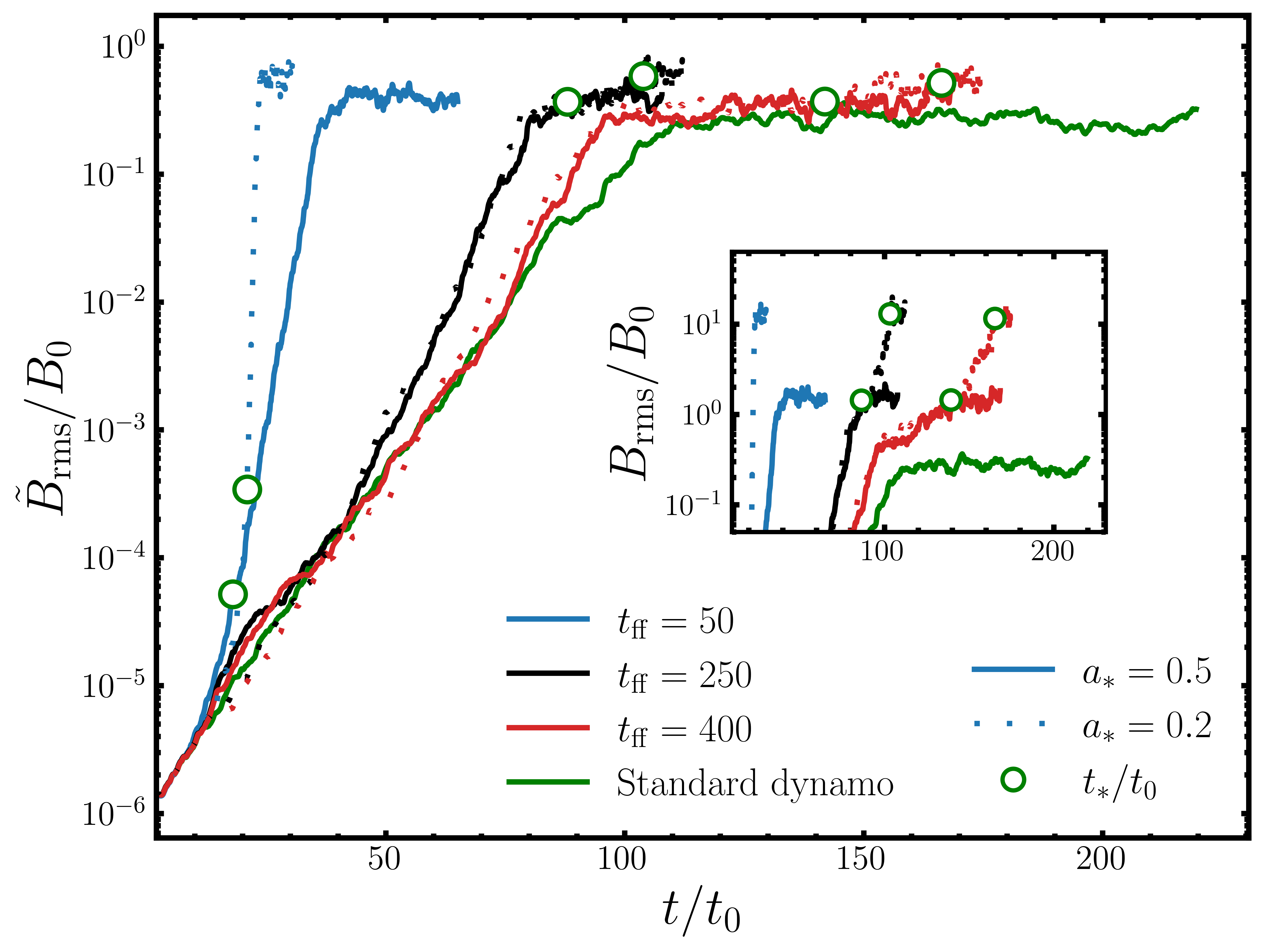}
	\caption{
		The rms supercomoving magnetic field compensated for the compression, $\wt{B}_\text{rms}/B_0$, in the SSD, for various values of $t_{\rm ff}$ {specified in the legend}. The solid and dotted curves represent two choices of the scale factor at which the collapse ends, $a\fr=0.5$ and 0.2, respectively, {and the green circles mark the corresponding time}. The inset presents the physical $B_{\rm rms}$ focusing on the nonlinear stage.
		The blue curves correspond to $\tff < t\nl$ and the others to $\tff \gtrsim t\nl$. {{T}he green curve{s} {represent} the dynamo in {a} stationary background ($a=1$).}
	}
	\label{fig:collapse_ssd5}
\end{figure}

Figure~\ref{fig:collapse_ssd5} and {Fig.}~\ref{fig:collapse_lsd5} {in the end matter} show the evolution of magnetic field in the SSD and LSD, respectively. 
The main frames isolate the effect of the dynamo action on the supercomoving rms magnetic field strength (presenting $\wt{B}_\text{rms}=a^2 B_\text{rms}$) whereas the insets, focusing on the nonlinear regime, show how the physical rms magnetic field $B_\text{rms}$ grows due to both compression and dynamo action. As Eqs~\eqref{eq:sim6} and \eqref{eq:kcoldynamo3} show, when $t_{\rm ff}$ is smaller 
(the cloud is denser), $\wt{a}$ {declines rapidly,} {yielding faster} magnetic field grow{th} in {the} kinematic stage. 
An important observation is the distinction in magnetic field evolution between the cases where $\tff < t\nl$ and $\tff \gtrsim t\nl$. To appreciate this, we note that in the stationary background {(green curve)} the kinematic stage ends (and the magnetic field strength levels off) 
in the SSD and LSD, 
significantly later than in a collapsing cloud with $t_\text{ff} < t\nl$. As a result, denser clouds 
develop strong magnetic fields earlier. Furthermore, if the dynamos were inefficient in a stationary background, the collapse would help it to reach the end of the kinematic regime \textit{much} faster. This kind of speed-up is less significant when $\tff \gtrsim t\nl$.

\begin{figure}
	\centering
	\includegraphics[width=3.4in, height=2.2in, keepaspectratio]{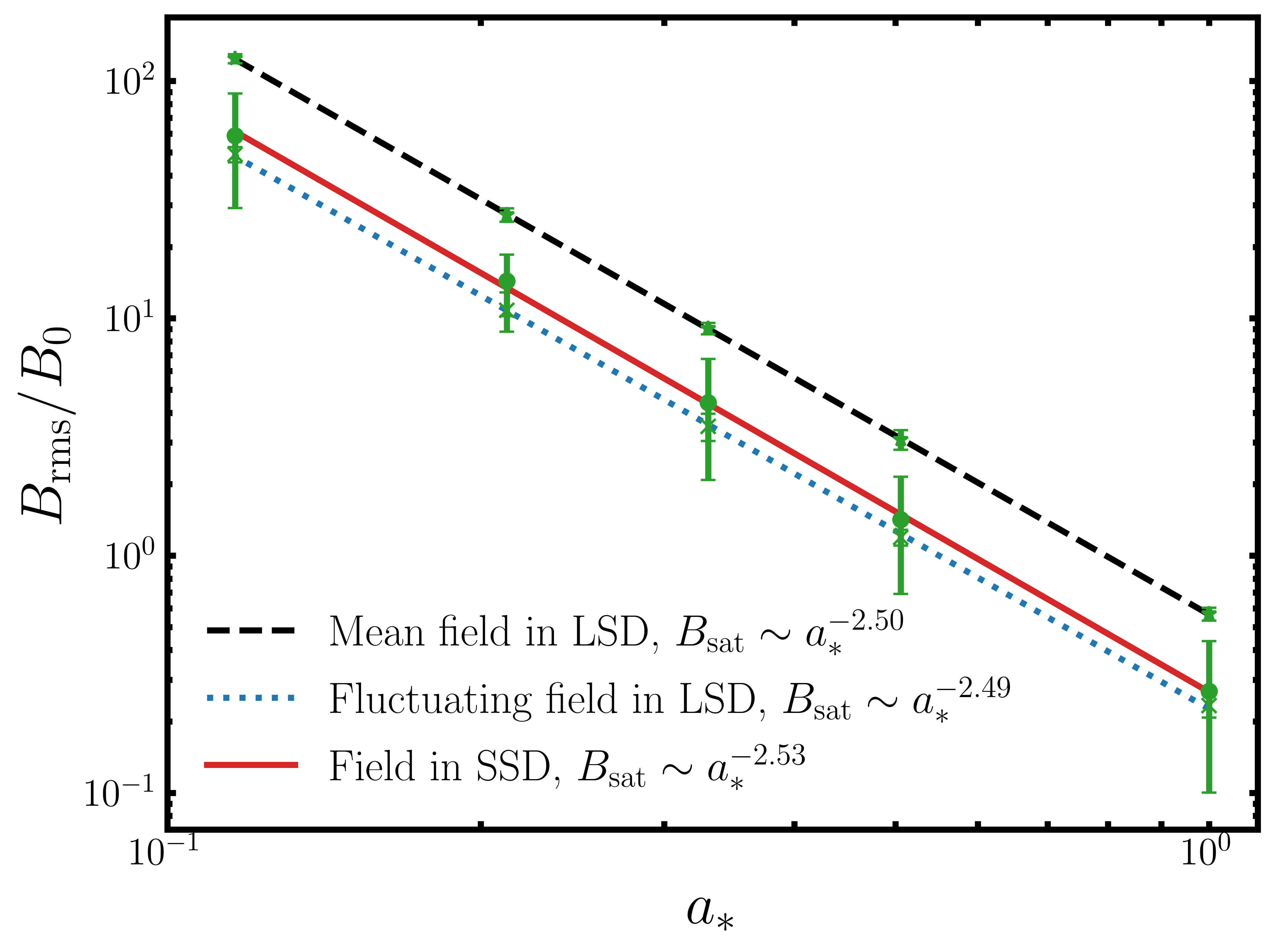}
	\caption{The dependence of the saturated rms strength of the physical magnetic field on $a\fr$ for $t_{\rm ff} = 50$. The error bars show the $5 \sigma$ deviation. The dashed lines show the scaling of Eq.~\eqref{eq:kcoldynamo4}. 
	}
	\label{fig:collapse_ssd_scaling}
\end{figure}

The supercomoving magnetic field continues to grow in the nonlinear regime in line with Eq.~\eqref{eq:kcoldynamo4}
which shows that $\tilde{B} \propto \tilde{a}(\wt{t})^{-1/2}$ arising from the weakening of supercomoving Lorentz force. The physical field
also grows due to the flux freezing as shown in  Eq.~\eqref{eq:kcoldynamo5}.
Despite a strong effect of the collapse 
on the kinematic 
dynamo, the saturated magnetic field strength is virtually independent of $\tff$ but is sensitive to the degree of compression, $a\fr$.
Figure~\ref{fig:collapse_ssd_scaling} shows the saturated physical field strength as a function of  $a\fr$.
In a remarkable agreement with Eq.~\eqref{eq:kcoldynamo5}, $B_{\rm rms}$ scales as $a\fr^{-5/2}$. We note that the magnetic field $\vect{B}$ is random with vanishing mean value in the case of the SSD and has a significant mean component in the LSD. As shown in Fig.~\ref{fig:collapse_ssd_scaling}, both parts of the magnetic field in the LSD scale similarly with $a\fr$.

\textit{Conclusions and discussion.---}
Dynamo action in a stationary environment exponentially amplif{y}  a seed magnetic field. We have shown that in a collapsing cloud the dynamo leads to a faster, super-exponential growth of the magnetic field. This effect is stronger than just a superposition of compression and exponential growth: the variable $a^2B$, where the overall impact of compression is factored out, also grows super-exponentially because the instantaneous growth rate of the magnetic field increases in the course of the collapse. 
The analysis in this work has benefited from using a framework of supercomoving variables that allowed us to recover the MHD equations in the collapsing background in nearly their original form.

{In general, local vortical motions in turbulent 
	flows are 
	amplified during {the} 
	collapse (because of the specific angular momentum conservation), leading to {a shorter} eddy turnover time and larger dynamo growth rate {as the collapse progresses}. 
	The results obtained here
	in the context of homologous collapse 
	can be adapted to any mode of the collapse (e.g., the hierarchical collapse \citep{V-sPB-PGZ19,TJMMLGN24}). In particular, the conclusion that dynamo action leads to a super-exponential amplification of the magnetic field applies to any collapsing system.}
{
	{As shown by us in \citep{SM}, super-exponential magnetic field amplification is possible even when the turbulence is not continuously driven but rather decays.} Motivated  by our work and using 
	the formalism developed here, \cite{2025AxelEvaDecay} {have} addressed this possibility, and confirmed it via simulations.
}

Apart from {enhancing} {the} {growth rate}, collapse also {increases} the steady-state 
strength of the magnetic field{.} 
For a spherically-symmetric collapse, a frozen-in magnetic field increases with the gas density $\rho$ as $B\propto\rho^{2/3}$, whereas the dynamo action leads to $B\propto\rho^{5/6}$ {in the continuously driven case}. {A steeper scaling with exponent between $2/3$ and $5/6$ is predicted even if the supercomoving turbulence is decaying \citep{SM}.}
Galaxies form from the intergalactic medium with a density enhancement by a factor of $10^5$--$10^6$.
Without any dynamo action, a frozen-in magnetic field would evolve from an initial value $B_i$ to 
$(\rho / \rho_0)^{2/3} B_i \simeq 10^4 B_i$.
However, with a dynamo operating, 
the final field strength,  
in equipartition with the turbulent flow, is  
$10^5 (4\pi \rho_0 \tilde v_0^2)^{1/2} $, where $(\rho / \rho_0)^{5/6} \simeq 10^5$. This is significantly stronger than what is achieved by flux-freezing alone or by a standard dynamo without collapse. 
{T}he contributions {of} the dynamo {action} and {compression with} flux-freezing  ({in addition to} the dynamo {amplification}) {can easily be separated in the framework of super-comoving coordinates}{.}
We discussed th{is} {for random (SSD) and large-scale (LSD) magnetic fields}{;}
{however, the analysis applies to any fast dynamo.}

The super-exponential growth of both random and mean magnetic fields in a collapsing turbulent cloud could 
significantly alter estimates of the age at which young galaxies develop observable magnetic fields. 
Current dynamo models 
which do not include this effect explain 
large-scale magnetic fields at the redshift $z\simeq3$ \citep{RCSBT19,JCSSRB24}, whereas random magnetic fields can become significant even earlier. 
Ordered galactic magnetic fields have been reported at redshifts up to $z=2.6$ \citep{RVPNPM24} and even $z=5.6$ \citep{chen_kiloparsec-scale_2024}. The polarized dust emission detected 
can emerge from both anisotropic random magnetic fields (which naturally occur in spiral arms 
due to differential rotation \citep{shukurovkandubook}) and mean magnetic fields. 
Whatever is the nature of the galactic magnetic fields at high redshifts, its super-exponential amplification may be a crucial aspect of the theory. 
To illustrate the new opportunities, we note that, in  
the case $\tff \sim 0.1t\nl$,
where $t\nl$ 
marks field saturation at presumably an observable strength,
the super-exponential growth over $15$ orders of magnitude 
reduces $t\nl$ of the standard dynamo by a factor of $10$ \citep{SM},
significantly affecting our understanding of galactic magnetic fields at high redshifts \citep{Heald2020}.
Cosmological MHD simulations of galaxy formation suggest turbulent dynamo action \citep{martin-alvarez18, pakmor24}, but evidence for super-exponential growth—potentially a key dynamo signature in a collapsing environment—has yet to be presented.

Our results also apply 
to primordial star formation{.}
{The efficient dynamo action resulting from the strong coupling between primordial gas and magnetic fields \citep{Maki2004, Maki2007, Nakauchi2019}, combined with the enormous compression factor from the molecular cloud ({density} $n\sim 10^{3} \cm^{-3}$) to the protostellar core 
	($n \sim 10^{24} \cm^{-3}$), 
	leads to significantly amplified magnetic fields in protostars \citep{sur_generation_2010}.}
However, it is {also} possible that the dynamo is  inefficient in these environments due to a low Prandtl number {or weak turbulence}, leading to a very large $t\nl$. Even then the 
collapse can significantly enhance the dynamo, with the extent of the speed-up depending on how large the ratio $t\nl/\tff$ is. A more detailed calculation of both the speed-up timescales and saturated magnetic field strengths in early stars and galaxies will be presented in a future work.

{Our analysis {focuses} on homogeneous, incompressible supercomoving turbulence in {a} collapsing cloud, both decaying and {maintained by an explicit forcing,} {using a} formalism {which} can be extended to compressible flows. The {form of the} forcing used reproduces the amplification {of random flows} in {the} gravity-driven {turbulence} \cite{Higashi_2021} {although} the effects {of the density} inhomogeneities remain to be explored.}

\nocite{Davidson2015} 
{
	\textit{Acknowledgments.---}
}
\label{sec:acknowledgments}
{
	We thank Godwin Martin and the members of the ICTS
	{p}lasma {a}strophysics group, particularly Vinay Kumar, for
	useful discussions. This research was supported by the Department of Atomic Energy, Government of India, under Project
	No. RTI4001. All simulations were performed on the Contra
	computing cluster at the International Centre for Theoretical
	Sciences, Tata Institute of Fundamental Research.}
{
	We thank the referees for very positive and helpful comments that 
	{have} led to several improvements in the paper, in particular, 
	{regarding the}  dynamo action in decaying turbulence. M.I.P. acknowledges the warm hospitality of
	Inter-University Centre for Astronomy and Astrophysics
	(IUCAA).
}

{
	\textit{Data availability.---}
}
\label{sec:data}
The RMS magnetic field data that support the findings of this article are openly available~\cite{Data}. The spectral data is storage heavy and can be provided upon reasonable request.




\bibliographystyle{apsrev4-2} 
\bibliography{refsPRL1}


{
	\onecolumngrid
	\begin{center}
		\textbf{\large End Matter}
	\end{center}
}
{
	Here we present the LSD figures in a collapsing background, analogous to Figs.~\ref{fig:compare_ssd} and \ref{fig:collapse_ssd5} for the SSD in the main text; all other details are provided there.
}

\begin{figure}
	\centering
	\begin{minipage}{0.48\textwidth}
		\centering
		\includegraphics[width=3.4in, height=2.2in, keepaspectratio]{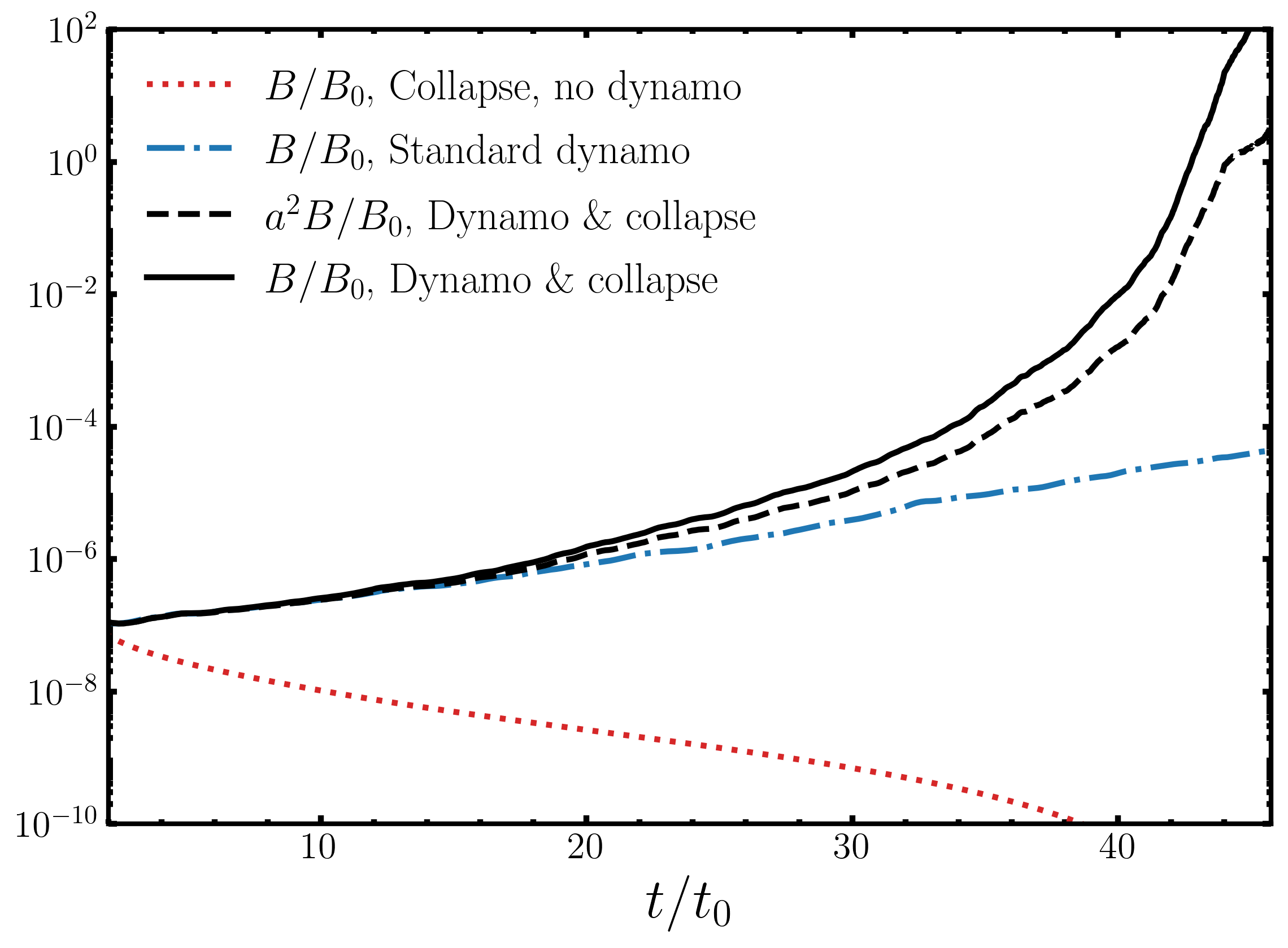}
		\caption{As Fig.~\ref{fig:compare_ssd} but for the LSD with $\Rey = 180$ and $\Rm = 18$.} 
		\label{fig:compare_lsd}
	\end{minipage}\hfill
	\begin{minipage}{0.48\textwidth}
		\centering
		\includegraphics[width=3.4in, height=2.2in, keepaspectratio]{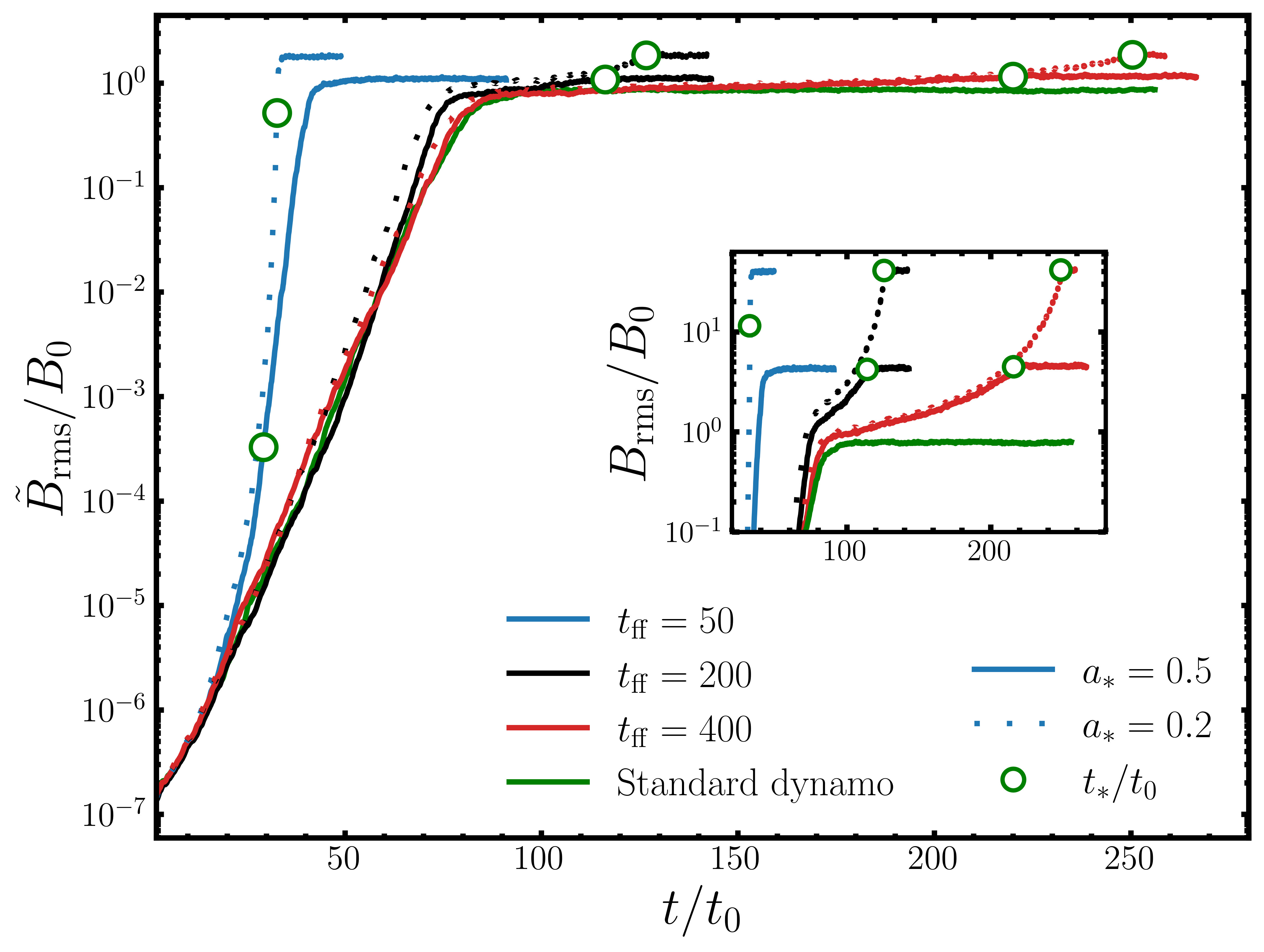}
		\caption{As Fig.~\ref{fig:collapse_ssd5} but for the LSD.}
		\label{fig:collapse_lsd5}
	\end{minipage}
\end{figure}

\clearpage
\onecolumngrid 
\begin{center}
\textbf{\large Supplemental Material}
\end{center}
\twocolumngrid 


\section{Dynamo equations and their solutions in supercomoving variables}
The interaction between a collisional plasma flow and magnetic field is described by the equations of magnetohydrodynamics (MHD),
\begin{gather}
\frac{\p\rho}{\p t} + \nabla\cdot\left(\rho\vect{u}\right) = 0,\label{eq:mhd1a}\\
\frac{\text{D} \vect{u} }{\text{D} t} 
= -\nabla \Phi -\frac{\nabla p}{\rho} + \frac{\left(\nabla\times\vect{B}\right)\times\vect{B}}{4\pi\rho}
+ \nu \nabla^2 \vect{u},\label{eq:mhd1b}\\
\frac{\p \vect{B}}{\p t} = \nabla\times\left(\vect{u}\times \vect{B}\right) + \eta \nabla^{2}\vect{B}, \quad \nabla\cdot\vect{B} = 0, 
\label{eq:mhd1c}
\end{gather}
where $\text{D}/\text{D}t= \partial/\partial t + (\vect{u}\cdot\nabla),$
$\vect{u}$, $p$, and $\rho$ are the gas velocity, pressure and density, respectively, $\Phi$ is the gravitational potential, $\vect{B}$ is the magnetic field, $\nu$ and $\eta$ are the kinematic viscosity and magnetic diffusivity, respectively.
The equations are closed by relating the pressure and density with the help of an equation of state.

As shown in the main text, the induction equation does not change its algebraic form when written in supercomoving variables. In this section, we present its solutions in supercomoving variables.

\subsection{The small-scale dynamo (SSD)}
Kazantsev's solution of the induction equation for the correlation function of the magnetic field in a random flow remains one of the most often used models of the SSD in a stationary background \citep{Kazantsev, shukurovkandubook}. The model assumes incompressible, statistically isotropic and
homogeneous, mirror-symmetric, $\updelta$-correlated in time Gaussian random velocity field. Consider such an incompressible  supercomoving velocity field $\wt{\vect{v}}(\wt{\vect{x}},\wt{t})$ and its supercomoving correlation tensor $\wt{T}_{ij}(\wt{r})$,
\begin{equation}\label{eq:supkaz2}
\left<\wt{v}_{i}(\wt{\vect{x}},\wt t)\; \wt{v}_{j}(\wt{\vect{y}},\wt s)\right> = \wt{T}_{ij}(\wt r)\,\updelta(\wt t-\wt s)\,,
\end{equation}
where $\wt{r}=|\wt{\vect{x}}-\wt{\vect{y}}|$ and we note that the velocity field is $\updelta$-correlated in both original and supercomoving variables. Since the continuity equation preserves its form when written in the supercomoving variables (Sect.~4.2 of Ref.~\citep{shapiro97}) and the gas density is assumed to be independent of position, $\wt{\nabla}\cdot\wt{\vect{v}}=0$ implies that the flow is incompressible in the physical variables as well, $\nabla\cdot\vect{v}=0$.
The velocity correlation tensor can be represented in terms of the longitudinal and transverse parts, $\wt{T}_\text{L}$ and $\wt{T}_\text{N}$, respectively, as $\wt{T}_{ij}=(\updelta_{ij}-\wt{r}_i\wt{r}_j/\wt{r}^2) \wt{T}_\text{N} + \wt{r}_i\wt{r}_j \wt{T}_\text{L}/\wt{r}^2$, where $\wt{T}_\text{L}$ and $\wt{T}_\text{N}$ are related to each other because of the flow incompressibility, similarly to the relation \eqref{MLMN} for the magnetic field correlators.

To derive the correlation tensor in terms of the peculiar velocity, we use
$\wt{\vect{v}}= a \vect{v}$ and $\wt t = f(t)$ in Eq.~\eqref{eq:supkaz2}:
\begin{equation}\label{eq:phdelta1}
\left<a(t) v_i(t)\; a(s) v_j(s)\right> = \wt{T}_{ij}(\wt r) \updelta\left(f(t) - f(s)\right).
\end{equation}
Since 
\[
\updelta[g(t)] = \frac{\updelta\left(t-t_0\right)}{|g'(t_0)|},
\]
where $t_0$ is the root of $g(t)$ (i.e., $t=s$ in our case), and
$\dd \wt t = \dd t/a^2$, we obtain
\[
\updelta\left[f(t) - f(s)\right] = \frac{1}{f'(t)|_{t=s}}\updelta(t-s) = a(s)^2 \updelta(t-s),
\]
implying that 
\begin{equation}\label{eq:phdelta2}
\left<v_i(t) v_j(s)\right> = \wt{T}_{ij}(\wt r) \updelta\left(t -s\right).
\end{equation}
Hence the peculiar velocity correlation tensor is $\updelta$-correlated in real time.

The magnetic field correlation tensor $\wt{M}_{ij}(\wt r, \wt t)$ cannot be assumed to be $\updelta$-correlated in time \cite{shukurovkandubook}, and it has the form
\begin{equation}\label{eq:supkaz3}
\left<\wt{B}_{i}(\wt{\vect{x}},\wt t)\;\wt{B}_{j}(\wt{\vect{y}},\wt t)\right> = \wt{M}_{ij}(\wt r, \wt t).
\end{equation} 
Since $\wt\nabla\cdot\wt{\vect{B}}=0$, the magnetic field correlation tensor $\wt{M}_{ij}$ can be written in terms of its longitudinal and transverse 
correlators
$\wt{M}_\text{L}$ and $\wt{M}_\text{N}$, $\wt{M}_{ij}=(\updelta_{ij}-\wt{r}_i\wt{r}_j/\wt{r}^2) \wt{M}_\text{N} + \wt{r}_i\wt{r}_j \wt{M}_\text{L}/\wt{r}^2$, which are related as
\begin{equation}\label{MLMN}
\wt{M}_\text{N} = \frac{1}{2\wt r}\frac{\p}{\p \wt r}\left(\wt{r}^2 \wt{M}_\text{L}\right).
\end{equation}

Since the supercomoving equations are similar to MHD equations in a stationary background, 
the 
derivation of the equation for the longitudinal magnetic correlator in Ref.~\citep{Kazantsev} remains applicable, and we obtain 
\begin{equation}\label{eq:supkaz4}
\frac{\p \wt{M}_\text{L}}{\p \wt{t}} = \frac{2}{\wt{r}^{4}}\frac{\p}{\p \wt{r}} \left(\wt{r}^{4}\wt{\eta}_\text{T}\frac{\p \wt{M}_\text{L}}{\p \wt{r}}\right) + \wt{G}\wt{M}_\text{L},
\end{equation}
where 
\[
\wt G = -4\left[\oderiv{}{\wt r}\left(\frac{\wt{T}_\text{N}}{\wt r}\right) + \frac{1}{\wt{r}^{2}}\oderiv{}{\wt r}\left(\wt{r}\wt{T}_\text{L}\right)\right]
\]
is responsible for the magnetic field amplification and $\wt{\eta}_\text{T} = \eta + \wt{T}_\text{L}(0) - \wt{T}_\text{L}(\wt r)$ represents the dissipation due to the electric resistivity and turbulent diffusion. This equation can be reduced to a Schr\"odinger-type equation and {solved using}
the WKBJ approximation{. For a Kolmogorov turbulent spectra, the magnetic energy amplifies in the initial kinematic stage as} 
{\citep{Kandu1997GR,BRANDENBURG20051,shukurovkandubook}}
\begin{equation}\label{eq:supkaz5}
\wt{\text B}_\text{rms} \propto \exp{\left(\wt\gamma \wt t \right)},\quad 
\wt\gamma \simeq {\frac{\wt{v}_\nu}{\wt{\ell}_\nu} = } \frac{\wt{v}_0}{\wt{l}_0} {\wt{\Rey}^{1/2}},
\end{equation} 
where $\wt{B}_\text{rms}$ is the rms magnetic field strength, {$\ell_\nu$, $v_\nu$ are the viscous eddy scale, velocity,} {and} ${{\wt{\Rey}}} = {\wt{v}_0\wt{l}_0 / \nu}$ {with} $\wt{v}_0$ and $\wt{l}_0$ are the integral speed and scale of the flow. Since $\wt{v}_0$ and $\wt{l}_0$ remain constant during the collapse, $\wt{\gamma}=\text{const}.$
{The initial kinematic stage continues until the magnetic energy reaches equipartition with the kinetic energy of the viscous eddies. After that the growth rate is governed by the next larger eddy and it continues until magnetic energy reaches equipartition with the integral scale over a time scale of the largest eddy turnover time.}

{While the $\delta$-correlation assumption simplifies our analysis, it is not essential. In standard dynamos, finite correlation times modify growth rates {to some degree} but preserve exponential growth \citep{Bhat2014, Gopalakrishnan2024}; we thus expect the key finding of super-exponential growth to remain robust.}

\subsection{The large-scale dynamo (LSD)}
The averaged supercomoving induction equation for the mean magnetic field $\mean{\wt{\vect{B}}}$ is given by
\begin{align}\label{eq:stdlsd1}
\frac{\p \mean{\wt{\vect{B}}}}{\p \wt t} &= \wt\nabla\times \left(\mean{\wt{\vect{V}}}\times \mean{\wt{\vect{B}}} + \wt{\bm{\mathcal{E}}} - \eta \wt\nabla\times\mean{\wt{\vect{B}}}\right),
\end{align}
where $\wt{\bm{\mathcal{E}}}=\mean{\wt{\vect{v}}\times \wt{\vect{b}}}$ and angular brackets denote a suitable averaging (e.g., ensemble or volume averaging or a filtering -- see Sect.~7.2 of Ref.~\cite{shukurovkandubook}).
It can be shown that $\wt{\bm{\mathcal E}}\simeq \wt{\alpha}\wt{\vect{B}} - \tilde\eta_\text{t} \wt{\nabla}\times\wt{\vect{B}}$ \cite{shukurovkandubook}, 
where $\tilde\alpha$ and $\tilde\eta_t$ are the turbulent transport coefficients 
\begin{align}\label{eq:stdlsd4}
\tilde\alpha \approx -\frac{1}{3} \tilde\tau_0 \left<\tilde{\vect{v}} \cdot \tilde{\bm{\omega}}\right>,\quad \tilde\eta_t \approx \frac{1}{3} \tilde\tau_0 \left<\tilde{\vect{v}}^2\right>,
\end{align}
where $\wt{\bm{\omega}}=\wt{\nabla}\times\wt{\vect{v}}$ and $\tau_0$ is the correlation time of the random flow. {The derivation assumes a small $\widetilde{\tau}_0$ and 
	$\Rm \gg 1$, such that the resistive and nonlinear terms in the fluctuations are {negligible in comparison} to the time derivative term.}
Assuming for simplicity that $\mean{\vect{V}}$ represents a solid-body rotation, the averaged induction equation written in the rotating frame follows as
\begin{align}\label{eq:stdlsd3}
\frac{\p \mean{\wt{\vect{B}}}}{\p \wt t} = \wt\nabla\times\left[\tilde\alpha \mean{\wt{\vect{B}}} \;
{-}\;
\wt{\eta}_\text{T} \wt\nabla\times\mean{\wt{\vect{B}}} \right],
\end{align}
with $\tilde\eta_\text{T}$ introduced in Eq.~\eqref{eq:supkaz4}. 
This type of the turbulent dynamo is known as the $\alpha^2$-dynamo.
The dependence of the growth rate of the magnetic field on the parameters can be illustrated using the simplest solution of Eq.~\eqref{eq:stdlsd3} in infinite space with $\wt{\alpha}$ and $\wt{\eta}_\text{T}$ independent of time and position (Section~7.5 of Ref.~\cite{shukurovkandubook} and Ref.~\cite{Sokoloff}),
\[
\wt\gamma =  \wt k \tilde\alpha -\wt{\eta}_\text{T} \wt{k}^2, 
\]
where $\wt{k}$ is the wavenumber of the mean magnetic field. A similar expression is valid for a spherically symmetric distribution of $\wt{\alpha}$ \cite{Sokoloff}. The magnetic field of the scale $2\pi/\wt{k}_\text{m}=4\pi\wt{\eta}_\text{T}/\wt{\alpha}$ grows most rapidly as $\exp(\wt{\gamma}_\text{m}\wt{t})$ with $\wt{\gamma}_\text{m}=\wt{\gamma}(\wt{k}_\text{m})=\wt{\alpha}^2/(4\wt{\eta}_\text{T})$.

Since $\wt t =f(t)$, the magnetic field grows super-exponentially in the physical time. It can be argued that the magnetic field growth rate in physical variables $\gamma_\text{m}=\alpha^2/(4\eta_\text{T})$ increases as $a$ decreases but the growth rate in the supercomoving variables $\wt{\gamma}_\text{m}$ remains constant. The supercomoving scale and rms speed of the random flow remain constant during the collapse, $\wt{l}_0=l_0/a=\text{const}$ and $\wt{v}_0=a v_0=\text{const}$, whereas the correlation time varies as $\tau_0\simeq l_0/v_0\propto a^2$. This is true at the kinematic stage of the dynamo action since the Lorentz force is negligible and the remaining terms in the  supercomoving momentum equation do not include $a$ explicitly. As a result, the mean helicity density of the random flow $\langle\vect{v}\cdot\bm{\omega}\rangle$ increases during the collapse as $\langle\vect{v}\cdot\bm{\omega}\rangle\propto a^{-3}$, while the turbulent magnetic diffusivity remain constant, $\eta_\text{t}=\text{const}$. This implies that the growth rate of the physical mean magnetic field varies as $\gamma_\text{m}\propto a^{-2}$ (we note that $\eta\ll\eta_\text{t}$ under normal conditions, so $\eta_\text{T}\approx \eta_\text{t}$). Since $\dd\wt{t}=\dd t/a^2$, this implies $\wt{\gamma}_\text{m}=\text{const}$.

\section{Duration of the kinematic stage}
During the kinematic regime of the SSD in a homologous collapsing background, 
the physical magnetic field evolves as 
\begin{equation}\label{eq:order1}
B = \frac{B_0}{a^2} \exp{\left[\tilde{\gamma}f(t)\right]},  
\end{equation}
where $B_0$ is the initial field, with the scale factor approximated with
\begin{equation}\label{eq:order}
a(t) = \left(1 - \frac{t^2}{\tff^2}\right)^{2/3},
\end{equation}
and
\begin{equation}\label{eq:order2}
f(t) = \frac{4t_{\rm ff}}{\pi} \sqrt{\frac{1-a}{a}},
\end{equation}
with $t_{\rm ff}$ the free-fall time and $\tilde{\gamma}$ the growth rate. 
The time at which kinematic regime ends for the
dynamo in a stationary background (the `standard dynamo') is related to $\wt{\gamma}$ and $B_0$ as
\begin{equation}\label{eq:order3}
t\nl^{\rm std} = \frac{1}{\tilde{\gamma}}\ln{\left(\frac{B\nl}{B_0}\right)}, 
\end{equation}
where $B\nl$ is the magnetic field strength at which the Lorentz force becomes sufficiently strong to affect the flow and the magnetic field growth slows down. Substituting Eqs~\eqref{eq:order3} and \eqref{eq:order2} in Eq.~\eqref{eq:order1}, we get
\begin{equation}\label{eq:order4}
\ln{\left(\frac{a^2 B\nl}{B_0}\right)} = \frac{4}{\pi}\frac{t_{\rm ff}}{t_{\rm nl}^{\rm std}} \sqrt{\frac{1-a}{a}} \;\ln{\left(\frac{B\nl}{B_0}\right)}.    
\end{equation}
To derive the scale factor, $a\nl$, at which kinematic regime ends for the dynamo in a collapsing background, 
we compare the left- and right-hand sides of
Eq.~\eqref{eq:order4} 
to obtain the corresponding time $t\nl$ using Eq.~\eqref{eq:order}. 

For example, consider the amplification of a seed magnetic field by 15 orders of magnitude, $B\nl / B_0 = 10^{15}$
in a cloud with $t_{\rm ff} = 0.1 t_{\rm nl}^{\rm std}$. From Eq.~\eqref{eq:order4}, we obtain 
$a\nl = 0.025$, corresponding to $t\nl = 0.09 t\nl^{\rm std},$ using Eq.~\eqref{eq:order}. 
Thus, it takes a ten times shorter time for a dynamo in the collapsing cloud to reach its nonlinear stage than in a stationary background.



{
	\section{{Dynamo action of} decaying turbulence}
}
{{Super-exponential amplification of magnetic field can occur even if turbulence in a collapsing cloud is not continuously driven but rather decays. Thus, we}
	consider the case of an initial{ly} turbulent plasma, with no forcing {and} a weak {seed} magnetic fields undergoing gravitational collapse. 
	{Now} we have decaying turbulence, unlike {the case of a driven turbulence explored} in the main text. 
}

{
	The supercomoving (SC) MHD equations in the kinematic stage ({when} the Lorentz force {is negligible}) are same as the standard MHD equations. Since the initial magnetic field is weak, {the turbulence decays as in a purely hydrodynamic case.} {In a Kolmogorov turbulence, the} rate of decay of supercomoving {kinetic} energy is {given by}
	\begin{equation}\label{eq:dec1}
	\frac{\dd \wt{v}^2}{\dd \wt t} = -{\etad}\; \frac{\wt{v}^3}{\wt l},
	\end{equation}
}
{where $\etad$ is the dissipation factor \cite{Robertson_2012}.} 

{
	From the conservation of {either the} angular momentum (Loitsyansky invariant) or {the} linear momentum (Saffman {invariant}), we have the following constraint on the {evolution of the} outer {turbulent scale:}
	\begin{equation}\label{eq:dec2}
	\wt{v}_0^\alpha \wt{l}_0 = \const = {\wt{v}_{0i}^\alpha \wt{l}_{0i} = \psi_0,}
	\end{equation}
}%
{
	where $\alpha = 2/5$ {if the} Loitsyansky {invariant applies} and $\alpha = 2/3$ {in the case of the} Saffman {invariant} \cite{Davidson2015},} 
{
	{a}nd the subscript $i$ denotes the initial values.}
{We assume a peaked velocity spectrum as is standard in decaying turbulence literature. Thus we can} {combine} { Eq.~\eqref{eq:dec2} with Eq.~\eqref{eq:dec1} {to obtain} $\wt{v}_0 \propto \wt{t}^{-{1} / {\left(1+\alpha\right)}}$.}
{Substituting {this} in Eq.~\eqref{eq:dec2} gives the evolution of {the} length scale as $\wt{l}_0 \propto \wt{t}^{{\alpha} / {\left(1 + \alpha\right)}}.$
}

{
	Magnetic field amplification is possible if the system is super-critical {with respect to the dynamo action}, that is magnetic Reynolds number is larger than the critical Reynolds number, $\Rm > \Rm^\text{cr}$. The amplification {continues as long as} {the eddies available in the spectrum remain} 
	{super}-critical. 
	During this period, the magnetic field {strength grows as}
	\begin{align}\label{eq:GR}
	a^2 B  = B_0 \exp{\left(\int \wt\gamma\, \dd\wt{t}\right)},
	\end{align}
}%
{
	where $B_0$ is the seed field {strength}, $a$ is the scale factor and $\wt\gamma$ is {the} turnover rate {(in the SC frame)} of the fastest eddy {available}. {As discussed in the main text, an} exponential growth in the SC coordinates (which is similar to the standard kinematic dynamo) {implies a} super-exponential amplification in the physical frame.}

{
	In {the} decaying turbulence, {the growth rate} $\wt\gamma_0$ {due to the largest} super-critical eddy of {a} scale $\wt{l}_0$ with $\Rm(\wt{l}_0) > \Rm^\text{cr}$ {is given by its} turnover rate,
	\begin{equation}
	\wt\gamma_0 = \frac{\wt{v}_0}{\wt{l}_0} \propto \frac{1}{\wt{t}},
	\end{equation}
}%
{
	and using the Kolmogorov scaling $\wt{v}(\wt{l}) \propto \wt{l}^{1/3}$ {we obtain} the growth rate due to eddy of scale $\wt l$
	\begin{align}
	\wt{\gamma}(\wt{l}) = \frac{\wt{v}}{\wt{l}} 
	= \frac{1}{\wt{l}}\frac{\wt{v}_0 \wt{l}^{1/3}}{\wt{l}_0^{1/3}}
	= \frac{\wt{v}_0}{\wt{l}_0} \left(\frac{\wt{l}_0}{\wt{l}}\right)^{2/3}.
	\end{align}
}


{
	Focusing on the stretching of field lines due to the eddy of the size $\wt{l}$, we fix $\wt{l}$ to obtain the growth rate as 
	\begin{align}
	\wt{\gamma}(\wt{l}) = \frac{\wt{v}_0}{\wt{l}_0^{1/3}} \frac{1}{\wt{l}^{2/3}} = \wt{\gamma}_C (\wt l)\,
	\wt{t}^{-(3 + \alpha)/[3(1+\alpha)]}.
	\end{align}
}
{
	where $\wt{\gamma}_C (\wt l)$ is a constant of proportionality.
}

{
	For the case of {the} Loitsyansky {invariant}, we obtain $\alpha = 2/5$, hence
	\begin{equation}
	\wt{\gamma}(\wt{l}) = \wt{\gamma}_C(\wt{l}) \wt{t}^{-{17}/{21}}.
	\end{equation}
}


{
	{Equation}~\eqref{eq:GR} {then shows that} {the} magnetic field {strength increases} as
	\begin{align}\label{LB}
	a^2 B &= B_0 \exp{\left(\wt{\gamma}_C(\wt l)\int \wt{t}^{-17/21}\dd\wt{t}\right)}, \notag
	\\
	&= B_0 \exp{\left(\frac{21}{4}\wt{\gamma}_C\wt{t}^{4/21}\right)}.
	\end{align}
}
{
	Similarly, for the Saffman-type {decay} we obtain $\wt{\gamma}(\wt{l}) = \wt{\gamma}_C(\wt{l}) \wt{t}^{-11/15}$ {and}}
\begin{align}\label{SB}
{a^2 B = B_0 \exp{\left(\frac{15}{4}\wt{\gamma}_C\wt{t}^{4/15}\right)}.}
\end{align}

{
	{The field strength in a freely decaying turbulence increases slower than in a steady supercomoving turbulence discussed in the main text, where the power of $\wt{t}$ in the exponent is unity.}  From the definition of $\wt t$ in Eq.~(3) and Eq.~(9) of the main text, it is clear that $\wt{t}$ is a steeply increasing function of $t$. {We conclude that a} super-exponential {magnetic field} growth {discussed} in the main text {occurs,} albeit with a different rate of growth, even in {this} least favorable scenario{, the case} of decaying turbulence. 
}

{
	We emphasize that {the} magnetic field {can only grow} in a super-critical decaying turbulence. Once the turbulence becomes subcritical, the magnetic field will decay. But as far as astrophysical systems are considered, $\Rm\gg \Rm^\text{cr}$, and {it} can take a long time (say, $t_R$) to reduce $\Rm$ to $\Rm^\text{cr}$, {thus} providing sufficient time for the magnetic field amplification. {When} $t_R \gtrsim \tff$, we get {a} significant {additional} super-exponential growth due to the collapse. {We also note that} maintaining $\Rm > \Rm^\text{cr}$ for {a} sufficient time {in numerical simulations and thus} {detecting} the scaling{s} {of Eqs~\eqref{LB} and/or \eqref{SB} can be difficult because of computational constraints on the numerical resolution.} 
}

{At the end of the kinematic stage, the back{-}reaction from the Lorentz force saturates the magnetic field growth. {T}he {dependence} of {the} magnetic field {strength on} {the gas} density in the {saturated dynamo can be derived as follows}.} 

{The {dependence of the flow speed on the} scale factor is obtained by integrating Eqs~\eqref{eq:dec1} and \eqref{eq:dec2},
	\begin{align}\label{eq:nldec1}
	\int_{\wt{v}_{0i}}^{\wt{v}_0} \wt{v}^{-(2+\alpha)} \dd \wt{v} &= -\frac{\eta_{\rm d}}{2\psi_0} \int_0^{\wt{t}}\dd \wt t', \notag\\
	\wt{v}_0^{-(1+\alpha)} &= \wt{v}_{0i}^{-(1+\alpha)} + \frac{\eta_{\rm d}(1+\alpha)}{2\psi_0} \wt{t}, \notag\\
	\wt{v}_0 &= \wt{v}_{0i} \left[1 + \frac{\eta_{\rm d}(1+\alpha)}{2\psi_0 \wt{v}_{0i}^{-(1+\alpha)}} \wt{t}\right]^{-1/(1+\alpha)}.
	\end{align}
	{From} $\psi_0 = \wt{v}_{0i}^{\alpha}\wt{l}_{0i}$, we {obtain} $\psi_0 \wt{v}_{0i}^{-(1+\alpha)} = \wt{l}_{0i} / \wt{v}_{0i} = \tau_L${, the turnover time of} the {outer} eddy. {Using}
	\[
	\wt{t} = \frac{2}{s}\sqrt{\frac{1-a}{a}}, \quad s = \frac{\pi}{2\tff}, 
	\]
	Eq.~\eqref{eq:nldec1} {can be written} in terms of {the} scale factor {as}
	\begin{align}\label{eq:nldec2}
	\wt{v}_0 &= \wt{v}_{0i} \left[1 + (1+\alpha)\chi \sqrt{\frac{1-a}{a}}\right]^{-1/(1+\alpha)},
	\end{align}
	where $\chi = \eta_{\rm d} / s \tau_L \sim \tff / t\nl.$}

{
	The saturation stage of the dynamo is characterized by the equipartition between magnetic and kinetic energies (Eq.~(10) of the main text),
	\begin{align}\label{eq:nldec3}
	a(t) \frac{\wt{B}^2}{8\pi} \simeq \frac{1}{2}\wt{\rho}\wt{v}_0^2.
	\end{align}
	Assuming {that} Eq.~\eqref{eq:nldec2} {applies} during the nonlinear dynamo stage, 
	we obtain {for $a\ll 1$}, the scaling
	\begin{align}\label{eq:nldec4}
	\wt{v}_0 \propto a^{1 / [2(1+\alpha)]} \propto \rho^{-1 / [6(1+\alpha)]}, 
	\end{align}
	{whereas} the peculiar velocity field scales as
	\begin{align}\label{eq:nldec5}
	v_0 \propto a^{-(1+2\alpha)/2(1+\alpha)} \propto \rho^{(1+2\alpha)/6(1+\alpha)}. 
	\end{align}
	Finally{,} from Eq.~\eqref{eq:nldec3}, we obtain the scaling of {the} magnetic field {strength as}
	\begin{align}\label{eq:nldec6}
	B \propto \sqrt{\rho v_0^2} \propto a^{-(4+5\alpha)/2(1+\alpha)} \propto \rho^{(4+5\alpha)/[6(1+\alpha)]}.
	\end{align}
	In the limit of $\alpha\to 0$, {this expression reproduces} the flux-freezing exponent of 2/3{,} and the exponent 5/6 {of} the forced case is retrieved in the {limit} $\alpha \to \infty$ which corresponds
	to {a} constant $\tilde{v}$. The exponent is greater than 2/3 for any $\alpha> 0$, irrespective{ly} of the invariant that governs the decay. After the collapse, the magnetic field can be sustained by other processes such as convection {in the case of stars,} {magnetic or fluid instabilities for} accretion-disk flows and stellar feedback in {the case of} galaxies.}


\section{The forcing function}
We use the \textsc{Pencil Code} forcing function $\wt{\vect{f}}$ written in supercomoving coordinates to generate a random flow in the simulations \citep{PCC21, Candelaresi_2013}, 
\begin{align}\label{eq:sim2}
\wt{\vect{f}}\left(\wt{\vect{x}}, \wt t\right) &= \mathrm{Re}\left\{N\vect{f}_{\wt{\vect{k}}(\wt t)}\; \exp{\left[\ii\wt{\vect{k}}(\wt t)\cdot\wt{\vect{x}} + \ii\phi(\wt t)\right]} \right\}, \\
N &= f_0 \wt{v}_{0} \left(\frac{\wt{k} \wt{v}_{0}}{\Delta \wt t}\right)^{1/2},\quad \vect{f}_{\wt{\vect{k}}} = \vect{R}\cdot \vect{f}_{\wt{\vect{k}}}^{(0)}\notag\\
R_{ij} &= \frac{\updelta_{ij} - \ii\sigma \epsilon_{ijk}\hat{\wt{k}}_k}{\sqrt{1 + \sigma^2}},\quad \vect{f}_{\wt{\vect{k}}}^{(0)} = \frac{\wt{\vect{k}}\times\vect{e}}{\sqrt{\wt{k}^{2} - (\wt{\vect{k}}\cdot\vect{e})^{2}}},\notag
\end{align}
where $\wt{\vect{k}}$ is the supercomoving wave-vector. The normalization factor $N$ decreases with the time step $\Delta\wt{t}$ to ensure 
that the forcing is approximately $\updelta$-correlated in the supercomoving time. 
The time step is calculated using the CFL condition. 
{Although the forcing function is $\delta$-correlated in $\tilde{t}$, the resulting supercomoving velocity is not. However, this forcing is sufficient to generate randomness in the flow, essential for exciting turbulent dynamos.}
The remaining factors in $N$ are fixed using dimensional arguments with a dimensionless parameter $f_0$ to control the forcing strength; 
we set $f_0 = 1$. The factor $\sigma$ in $R_{ij}$ is a measure of the kinetic helicity. For SSD we use $\sigma=0$ to
drive a non-helical flow, whereas $\sigma=1$ for LSD to generate a helical velocity field required to produce a large-scale magnetic field.

\section{Numerical resolution tests}
\begin{figure}[h!]
	\centering
	\includegraphics[width=3.4in, height=2.2in, keepaspectratio]{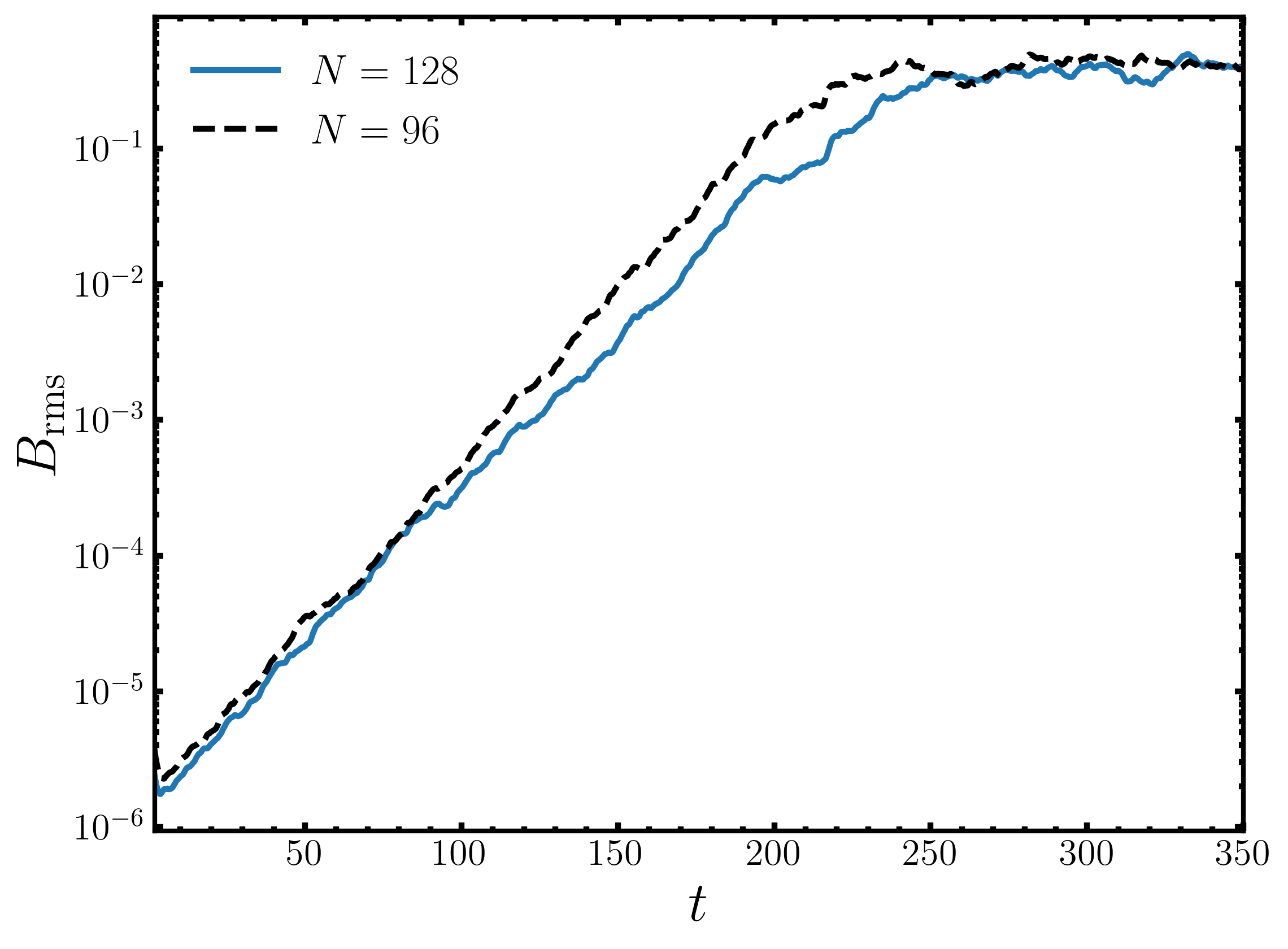}
	\caption{Comparison of the evolution of $B_{\rm rms}$ 
		in the SSD for the numerical resolutions $128^3$ and $96^3$. 
		{Parameter values are the same as in Figs~\ref{fig:compare_ssd} and \ref{fig:collapse_ssd5} of the main text but the results are presented here versus $t$ rather than $t/t_0$.}
	}
	\label{fig:ssd128-96comp}
\end{figure}

\begin{figure}[h!]
	\centering
	\includegraphics[width=3.4in, height=2.2in, keepaspectratio]{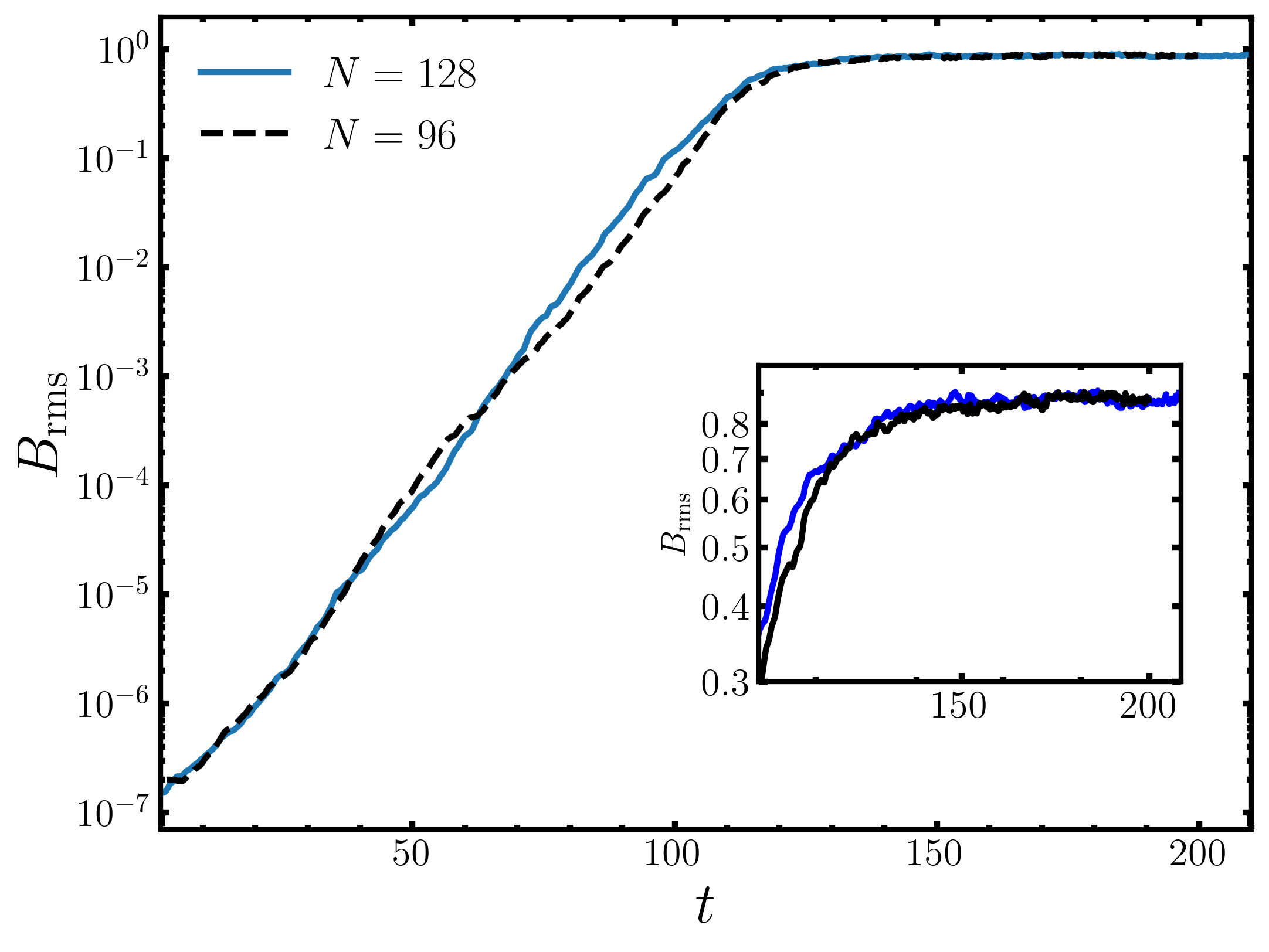}
	\caption{As Fig.~\ref{fig:ssd128-96comp} but for the LSD. The inset shows the slow transition to the dynamo saturation completing around ${t = 200}$.}
	\label{fig:lsd128-96comp}
\end{figure}

Here we verify that the numerical resolution of $128^3$ is sufficient for our purposes. Figures \ref{fig:ssd128-96comp} and \ref{fig:lsd128-96comp} show the evolution of the rms magnetic field strength $B_{\rm rms}$ in two simulations of SSD and LSD, respectively, with the resolutions $128^3$ and $96^3$. 
{The SSD simulation uses $k_f = 2, \Rey = 415$, and $\Pm = 1$, while the LSD simulation employs $k_f = 4, \Rey = 180$, and $\Pm = 0.1$.}
We find the agreement between the results obtained under the two numerical resolutions to be sufficiently close to justify $128^3$ as the resolution used to obtain our main results.


\end{document}